\documentstyle[preprint,aps]{revtex}
\tightenlines
\begin{document}
\title{ENERGY DISTRIBUTION IN MELVIN'S MAGNETIC UNIVERSE}
\author{S.~S.~Xulu\thanks{%
E-mail: ssxulu@pan.uzulu.ac.za}}
\address{Department of Applied Mathematics, University of Zululand,\\
Private Bag X1001, 3886 Kwa-Dlangezwa, South Africa}
\maketitle

\begin{abstract}
We use the energy-momentum complexes of Landau and Lifshitz and Papapetrou
to obtain the energy distribution in Melvin's magnetic universe. For this
space-time we find that these definitions of energy give the same and
convincing results. The energy distribution obtained here is the same as we
obtained earlier for the same space-time using the energy-momentum complex
of Einstein. These results uphold the usefulness of the energy-momentum
complexes.
\end{abstract}



\section{Introduction}

The well-known Melvin's magnetic universe\cite{Melvin1} is a collection of
parallel magnetic lines of force in equilibrium under their mutual
gravitational attraction. This magnetic universe is described by an
electrovac solution to the Einstein-Maxwell equations \footnote{%
This solution had been obtained earlier by Misra and Radhakrishna\cite{Misra}%
. Melvin\cite{Melvin2} mentioned that this solution is contained implicitly
as a special case among the solutions obtained by Misra and Radhakrishna.}.
This space-time is invariant under rotation about, and translation along, an
axis of symmetry. This is also invariant under reflection in planes
comprising that axis or perpendicular to it. Wheeler\cite{Wheel}
demonstrated that a magnetic universe could also be obtained in Newton's
theory of gravitation and showed that it is unstable according to elementary
Newtonian analysis. Further Melvin\cite{Melvin2} showed his universe to be
stable against small radial perturbations and Thorne\cite{Thorne} proved the
stability of the magnetic universe against arbitrary large perturbations.
Thorne\cite{Thorne} further pointed out that the Melvin magnetic universe
might be of great value in understanding the nature of extragalactic sources
of radio waves and thus the Melvin solution to the Einstein-Maxwell
equations is of immense astrophysical interest. Virbhadra and Prasanna\cite
{VirPra} studied spin dynamics of charged massive test particles in this
space-time. It is really tempting to investigate the energy distribution in
the Melvin magnetic universe.

The energy-momentum localization or quasi-localization in general relativity
remains an elusive problem. In a flat space-time the energy-momentum tensor $%
T_{i}^{\ k}$ satisfies the divergence relation $T_{i\ ,k}^{\ k}=0$. The
presence of gravitation necessitates the replacement of an ordinary
derivative by a covariant one, and therefore one has the covariant
conservation laws $T_{i\ ;\ k}^{\ k}=0$. In a curved space-time the
energy-momentum tensor of matter plus all non-gravitational fields does not
satisfy $T_{i\text{ },\text{ }k}^{\ k}=0$; the contribution from the
gravitational field is also required to construct an energy-momentum complex
which satisfies a divergence relation like one has in a flat space-time.
Attempts aimed at finding a quantity for describing distribution of
energy-momentum due to matter, non-gravitational and gravitational fields
resulted in various energy-momentum complexes, notably those proposed by
Einstein, Landau and Lifshitz, Papapetrou and Weinberg. The physical meaning
of these nontensorial (under general coordinate transformations) complexes
have been questioned by some researchers (see references in \cite{ChaFe91}).
There are suspicions that different energy-momentum complexes could give
different energy distributions in a given space-time. The problems
associated with energy-momentum complexes resulted in a number of
researchers questioning the concept of energy-momentum localization.
According to Misner {\em et al.} \cite{MTW} the energy is localizable only
for spherical systems. However, Cooperstock and Sarracino \cite{CoopSar}
refuted their viewpoint and stated that if the energy is localizable in
spherical systems then it is also localizable for all systems. Bondi\cite
{Bondi} wrote `` {\em In relativity a non-localizable form of energy is
inadmissible, because any form of energy contributes to gravitation and so
its location can in principle be found}.'' A large number of definitions of
quasi-local mass have been proposed (see in \cite{BroYor},\cite{Hayw}).
Bergqvist \cite{Bergq} furnished computations with many definitions of
quasi-local masses for the Reissner-Nordstr\"{o}m and Kerr space-times and
came to the conclusion that no two of these definitions gave the same result.

Virbhadra, his collaborators and some others\cite{VirAsy} considered many
asymptotically flat space-times and presented that several energy-momentum
complexes give the same and acceptable results for a given space-time. Rosen
and Virbhadra \cite{RosVir} investigated the Einstein-Rosen space-time
(which is not asymptotically flat) and noted that many energy-momentum
complexes give the same and persuading results for the energy and energy
current densities. Fascinated by this interesting work some researchers
studied few asymptotically non-flat space-times using different energy-
momentum complexes and obtained appealing results (see \cite{AsyNon}- \cite
{Xulu}). Aguirregabiria {\em et al.} \cite{AgChVir} proved that several
energy-momentum complexes ``coincide'' for any Kerr-Schild class metric.
Recently Virbhadra\cite{Virbh99} showed that for a general non-static
spherically symmetric metric of the Kerr-Schild class, the energy-momentum
complexes of Einstein, Landau and Lifshitz, Weinberg and Papapetrou furnish
the same result as Tod obtained using the Penrose quasi-local mass
definition. These are definitely inspiring results and it is worth pursuing
this research topic further. Recently there has been some other important
papers on this topic\cite{recent}.

Energy distribution in Melvin's universe was earlier computed \cite{Xulu}
using Einstein's complex. In this paper we obtain energy distribution in
Melvin's magnetic universe using the definitions of Landau and Lifshitz, and
Papapetrou. We wish to check whether or not we get the same result as we
obtained earlier using the energy-momentum complex of Einstein. In this
paper we use geometrized units where gravitational constant $G=1$ and the
speed of light in vacuum $c=1$. The convention used in this paper is that
Latin indices take values from 0 to 3 and Greek indices values from 1 to 3,
comma indicates ordinary derivative and semi-colon covariant derivative.

\section{Melvin's magnetic universe}

The Einstein-Maxwell equations are very well-known in the literature. These
are given by 
\begin{equation}
R_{i}^{\ k}-\frac{1}{2}\ g_{i}^{\ k}R=8\pi T_{i}^{\ k},
\end{equation}
\begin{equation}
\frac{1}{\sqrt{-g}}\left( \sqrt{-g}\ F^{ik}\right) _{,k}=4\pi J^{i}\ ,
\end{equation}
\begin{equation}
F_{ij,k}+F_{j,k,i}+F_{k,i,j}=0,
\end{equation}
where the energy-momentum tensor of the electromagnetic field is 
\begin{equation}
T_{i}^{\ k}=\frac{1}{4\pi }\left[ -F_{im}F^{km}+\frac{1}{4}\ g_{i}^{\
k}F_{mn}F^{mn}\right] .
\end{equation}
$J^{i}$ is the electric current density vector and $R_{i}^{\ k}$ is the
Ricci tensor. Melvin (see in \cite{Ernst}) obtained electrovac solution ($%
J^{i}=0$) to these equations, which is expressed by the line element 
\begin{equation}
ds^{2}=\Lambda ^{2}\left[ dt^{2}-dr^{2}-r^{2}d\theta ^{2}\right] -\Lambda
^{-2}r^{2}\sin ^{2}\theta d\phi ^{2}  \label{melvinle}
\end{equation}
and the Cartan components of the magnetic field 
\begin{eqnarray}
H_{r} &=&\Lambda ^{-2}B_{o}\cos \theta ,  \nonumber \\
H_{\theta } &=&-\Lambda ^{-2}B_{o}\sin \theta ,
\end{eqnarray}
where 
\begin{equation}
\Lambda =1+\frac{1}{4}B_{o}^{2}r^{2}\sin ^{2}\theta .  \label{Lambda}
\end{equation}
$B_{o}$ ($\equiv B_{o}\sqrt{G}/c^{2}$) is the magnetic field parameter and
this is a constant in the solution given above. The non-zero components of
the energy-momentum tensor are \cite{Xulu} 
\begin{eqnarray}
T_{1}^{\ 1} &=&-T_{2}^{\ 2}=\frac{B_{o}^{2}\left( 1-2\sin ^{2}\theta \right) 
}{8\pi \Lambda ^{4}},  \nonumber \\
T_{0}^{\ 0} &=&-T_{3}^{\ 3}=\frac{B_{o}^{2}}{8\pi \Lambda ^{4}},  \nonumber
\\
T_{2}^{\ 1} &=&-T_{1}^{\ 2}=\frac{2B_{o}^{2}\sin \theta \cos \theta }{8\pi
\Lambda ^{4}}.
\end{eqnarray}

To get meaningful results using these energy-momentum complexes one is
compelled to use ``Cartesian'' coordinates (see \cite{Moller} and \cite
{Virbh99}). Then the line element $(\ref{melvinle})$ is transformed to
``Cartesian'' coordinates $t,x,y,z$ using the standard transformation 
\begin{eqnarray}
r &=&\sqrt{x^{2}+y^{2}+z^{2}},  \nonumber \\
\theta &=&\cos ^{-1}\left( \frac{z}{\sqrt{x^{2}+y^{2}+z^{2}}}\right) , 
\nonumber \\
\phi &=&\tan ^{-1}(y/x).  \label{eqtn3}
\end{eqnarray}
Now, the line element in $t,x,y,z$ coordinates becomes 
\begin{equation}
ds^{2}=\Lambda ^{2}dt^{2}-\Lambda ^{2}(dx^{2}+dy^{2}+dz^{2})+\left( \Lambda
^{2}+\frac{1}{\Lambda ^{2}}\right) \frac{(xdy-ydx)^{2}}{x^{2}+y^{2}} \text{.}
\label{melvinlecart}
\end{equation}
The determinant of the metric tensor is given by 
\begin{equation}
g=-\Lambda ^{4}
\end{equation}
and the non-zero contravariant components of the metric tensor are 
\begin{eqnarray}
g^{_{00}} &=&\Lambda ^{-2},  \nonumber \\
g^{_{^{11}}} &=&-\,\frac{\Lambda ^{-2}x^{2}+\Lambda ^{2}y^{2}}{x^{2}+y^{2}},
\nonumber \\
g^{_{12}} &=&-\left( \Lambda ^{2}-\frac{1}{\Lambda ^{2}}\right) \frac{xy}{%
x^{2}+y^{2}},  \nonumber \\
g^{_{22}} &=&-\,\frac{\Lambda ^{-2}y^{2}+\Lambda ^{2}x^{2}}{x^{2}+y^{2}}, 
\nonumber \\
g^{_{33}} &=&-\Lambda ^{-2}.  \label{gupik}
\end{eqnarray}

\section{The Landau and Lifshitz energy-momentum complex}

The energy-momentum complex of Landau and Lifshitz\cite{LanLif} is 
\begin{equation}
L^{ij}=\frac{1}{16\pi }{\cal S}_{\quad ,kl}^{ikjl}  \label{Lij}
\end{equation}
where 
\begin{equation}
{\cal S}^{ikjl}=-g(g^{ij}g^{kl}-g^{il}g^{kj})  \label{Sijkl}
\end{equation}
$L^{ij}$ is symmetric in its indices. $L^{00}$ is the energy density and $%
L^{0\alpha }$ are the momentum (energy current) density components. $%
S^{mjnk} $ has symmetries of the Riemann curvature tensor. The expression 
\begin{equation}
P^{i}=\int \int \int L^{i0}dx^{1}dx^{2}dx^{3}  \label{eqtn9}
\end{equation}
gives the energy $P^{0}$ and the momentum $P^{\alpha }$ components. Thus the
energy $E$, after applying the Gauss theorem, is given by the expression 
\begin{equation}
E_{LL}=\frac{1}{16\pi }\int \int {\cal S}_{\quad ,\alpha }^{0\alpha 0\beta
}\ \mu _{\beta }\ dS  \label{LLGauss}
\end{equation}
where $\mu _{\beta }$ is the outward unit normal vector over an
infinitesimal surface element $dS$. In order to calculate the energy
component for Melvin's universe expressed by the line element $(\ref
{melvinlecart})$ we need the following non-zero components of ${\cal S}%
^{ikjl}$ 
\begin{eqnarray}
{\cal S}^{0101} &=&-\frac{x^{2}+y^{2}\Lambda ^{4}}{x^{2}+y^{2}}\text{,} 
\nonumber \\
{\cal S}^{0102} &=&\frac{xy(\Lambda ^{4}-1)}{x^{2}+y^{2}}\text{,}  \nonumber
\\
{\cal S}^{0202} &=&-\frac{y^{2}+x^{2}\Lambda ^{4}}{x^{2}+y^{2}}\text{,} 
\nonumber \\
{\cal S}^{0303} &=&-1\text{.}  \label{Scomponents}
\end{eqnarray}
Equation $(\ref{Lij})$ with equations $(\ref{Sijkl})$ and $(\ref{Scomponents}%
)$ gives 
\begin{equation}
L^{00}=\frac{1}{8\pi }B^{2}\Lambda ^{3}\text{.}  \label{L00}
\end{equation}
For a surface given by parametric equations $x=r\sin \theta \cos \phi ,$ $\
y=r\sin \theta \sin \phi ,$ $\ z=r\cos \theta $ (where $r$ is constant) one
has $\mu _{\beta }=\{x/r,$ $y/r,$ $z/r\}$ and $dS=r^{2}sin\theta d\theta
d\phi $. Using equations $(\ref{Scomponents})$ in $(\ref{LLGauss})$ over a
surface $r=const.$, we obtain 
\begin{equation}
E_{LL}=\frac{1}{6}B_{o}^{2}r^{3}+\frac{1}{20}B_{o}^{4}r^{5}+\frac{1}{140}%
B_{o}^{6}r^{7}+\frac{1}{2520}B_{o}^{8}r^{9}.  \label{ELL}
\end{equation}

\section{The Energy-momentum complex of Papapetrou}

The symmetric energy-momentum complex of Papapetrou\cite{Papp} is 
\begin{equation}
\Omega ^{ij}=\frac{1}{16\pi }{\cal {N}}_{\quad ,kl}^{ijkl}  \label{Omega}
\end{equation}
where 
\begin{equation}
{\cal {N}}^{ijkl}=\sqrt{-g}\left( g^{ij}\eta ^{kl}-g^{ik}\eta
^{jl}+g^{kl}\eta ^{ij}-g^{jl}\eta ^{ik}\right)  \label{Nijkl}
\end{equation}
and the Minkowski metric 
\[
\eta ^{ik}=\left( 
\begin{array}{cccc}
1 & 0 & 0 & 0 \\ 
0 & -1 & 0 & 0 \\ 
0 & 0 & -1 & 0 \\ 
0 & 0 & 0 & -1
\end{array}
\right) 
\]
$\Omega ^{00}$ \ and $\Omega ^{\alpha 0}$ \ are the energy and momentum
density components. \ The energy and momentum components are given by 
\begin{equation}
P^{i}=\int \int \int \Omega ^{i0}dx^{1}dx^{2}dx^{3}\text{.}  \label{eqtn17}
\end{equation}
Using the Gauss theorem, the energy $E_{P}$ for a stationary metric is thus
given by the expression 
\begin{equation}
E_{P}=\frac{1}{16\pi }\int \int {\cal N}_{\qquad ,\beta }^{00\alpha \beta }\
\ \mu _{\alpha }dS.  \label{PapGauss}
\end{equation}
To find the energy component of the line element $(\ref{melvinlecart})$, we
require the following non-zero components of $\ {\cal N}^{ijkl}$ 
\begin{eqnarray}
{\cal N}^{0011} &=&-(1+\frac{x^{2}+y^{2}\Lambda ^{4}}{x^{2}+y^{2}})\text{,} 
\nonumber \\
{\cal N}^{0012} &=&\frac{xy(\Lambda ^{4}-1)}{x^{2}+y^{2}}\text{,}  \nonumber
\\
{\cal N}^{0022} &=&-(1+\frac{y^{2}+x^{2}\Lambda ^{4}}{x^{2}+y^{2}})\text{,} 
\nonumber \\
{\cal N}^{0033} &=&-2\text{.}  \label{Ncomponents}
\end{eqnarray}

Equations $(\ref{Ncomponents})$ in equation $(\ref{Omega})$ give the energy
density component 
\begin{equation}
\Omega ^{00}=\frac{1}{8\pi }B^{2}\Lambda ^{3}\text{.}  \label{Omega00}
\end{equation}
Thus we find the same energy density as we obtained in the last Section, we
use Eq. $(\ref{Ncomponents})$ in $(\ref{PapGauss})$ over a 2-surface (as in
the last Section) and obtain 
\begin{equation}
E_{P}=\frac{1}{6}B_{o}^{2}r^{3}+\frac{1}{20}B_{o}^{4}r^{5}+\frac{1}{140}%
B_{o}^{6}r^{7}+\frac{1}{2520}B_{o}^{8}r^{9}.  \label{EPap}
\end{equation}
This result is expressed in geometrized units ($G=1$ and $c=1$). In the
following we restore $G$ and $c$ and get 
\begin{equation}
E_{P}=\frac{1}{6}B_{o}^{2}r^{3}+\frac{1}{20}\frac{G}{c^{4}}B_{o}^{4}r^{5}+%
\frac{1}{140}\frac{G^{2}}{c^{8}}B_{o}^{6}r^{7}+\frac{1}{2520}\frac{G^{3}}{%
c^{12}}B_{o}^{8}r^{9}.
\end{equation}
The first term $\frac{B_{o}^{2}r^{3}}{6}$ is the known classical value of
energy and the rest of the terms are general relativistic corrections. The
general relativistic terms increase the value of energy.


\section{Discussion and Summary}

The subject of energy-momentum localization in the general theory of
relativity has been very exciting and interesting; however, it has been
associated with some debate. Misner {\em et al.} \cite{MTW} argued that to
look for a local energy-momentum is looking for the right answer to the
wrong question. However, they further mentioned that energy is localizable
but only for spherical systems. Cooperstock and Sarracino\cite{CoopSar}
fully disagreed with them and argued that if the energy is localizable in
spherical systems then it is also localizable for all systems. Bondi\cite
{Bondi} wrote that a nonlocalizable form of energy is inadmissible in
general theory of relativity. There prevails scepticism that different
energy-momentum complexes could give unacceptable different energy
distributions for a given space-time. However, buttressed by the remarkable
results obtained by Virbhadra, his collaborators (Rosen, Parikh, Chamorro
and Aguirregabiria) and some others (who demonstrated with several examples
that for a particular space-time many energy-momentum complexes give the
same and acceptable energy distribution), this research topic is
rejuvenated. Recently Virbhadra\cite{Virbh99} stressed that although the
energy-momentum complexes are non-tensorial (under general coordinate
transformations), these do not violate the principle of general covariance
as the equations describing the conservation laws with these objects (for
example, $L_{\ ,k}^{ik}=0$, $\Omega _{\ ,k}^{ik}=0$) are true in any
coordinates systems.

In this paper we obtained the energy distribution in Melvin's magnetic
universe. We used the energy-momentum complexes of Landau and Lifshitz, and
Papapetrou. Both definitions give the same results ($L^{00}=\Omega
^{00},E_{LL}=E_{P}$). We also note that the results obtained here is the
same as we obtained earlier using the energy-momentum complex of Einstein.
The first term in the energy expression (see equations $(\ref{ELL})$ and $(%
\ref{EPap})$ is the well-known classical value for the energy of the uniform
magnetic field and the other terms are general relativistic corrections. The
general relativistic corrections increase the value of the energy. These
results uphold the importance of the energy-momentum complexes and oppose
the prevailing ``folklore'' against them.

\acknowledgments
I am grateful to K. S. Virbhadra for guidance, George F. R. Ellis for
hospitality at the university of Cape Town, and NRF (S. Africa) for
financial support.


\end{document}